\documentclass[fleqn,twoside]{article}
\usepackage{espcrc2}

\usepackage{graphicx}
\usepackage[figuresright]{rotating}

\newcommand{\AmS}{{\protect\the\textfont2
  A\kern-.1667em\lower.5ex\hbox{M}\kern-.125emS}}

\title{The Quenched Continuum Limit}

\author{C. T. H. Davies\address[GLA]{Department of Physics and Astronomy, 
         University of Glasgow, Glasgow, G12 8QQ, UK},
        G. P. Lepage\address{Laboratory of Elementary-Particle Physics,
                Cornell University, Ithaca NY 14850, USA},
        F. Niedermayer\address{ITP, University of Bern, Sidlerstrasse 5, CH-3012 Bern, Switzerland} 
        and
        D. Toussaint\address{Department of Physics, University of Arizona, 
                Tucson AZ 85721, USA.}}
       
\begin{document}

\begin{abstract}
We show that all current formalisms for quarks in lattice QCD are consistent
in the quenched continuum limit, as they should be. We improve on previous extrapolations to this limit, and the understanding of lattice systematic errors there, by using a constrained fit including both leading and sub-leading dependence on $a$. 
\end{abstract}

\maketitle

\section{INTRODUCTION}

Our increasing ability to carry out unquenched simulations means that 
the quenched approximation is no longer required. There is an outstanding
issue associated with the 
quenched continuum limit, however, which we 
address here. At the same time we point out the ingredients that are 
necessary to obtain an accurate result in the continuum limit, 
since these ingredients will also be important for unquenched simulations. 

The outstanding issue is the question of whether all quark formulations 
give the same result in the quenched continuum limit. It is relatively 
trivial to demonstrate that they should, because this is the limit in 
which it is easy to analyse the formulations. At Lattice 2000, however, it 
seemed not to be the case~\cite{aoki}. A plot was made of the ratio of 
nucleon to vector meson mass at a fixed physical quark mass, as a function 
of the lattice spacing. The physical quark mass was chosen as the point 
at which the pseudoscalar meson mass divided by the vector meson mass 
was 0.5 (in those days quite a low mass to reach). 
The lattice spacing was given in units of the vector meson mass. 
Because of the instability of vector mesons this is not a plot that allows
for a precision test against experimental results, even if it used a 
quark mass value from the real world and was not in 
the quenched approximation. 
However, it can be used as a comparison of different formalisms, provided 
the results have good statistical precision and are on large enough volumes.

The original plot showed a 6$\sigma$ disagreement
between the results extrapolated to $a=0$ from the Wilson formulation~\cite{CP-PACS} and 
the unimproved staggered formulation~\cite{MILC}. The Wilson results were 
extrapolated linearly in $a$ and the staggered results quadratically. 
Clover results were also included but their variation suggests a statistical 
error that means they are not particularly useful so we drop those 
results in this discussion. 

The disagreement, if true, would represent a major problem for lattice 
calculations and call the whole approach into question. With 
many large dynamical quark projects well underway and others planned, 
it is therefore time to readdress this issue, following on from~\cite{jansen}.  
In addition, there are more quark formulations in widespread use than there were 
in 2000, so the comparison of results can be significantly extended. 

\begin{figure}[htb]
\vspace{9pt}
\rotatebox{270}{\includegraphics[width=\hsize]{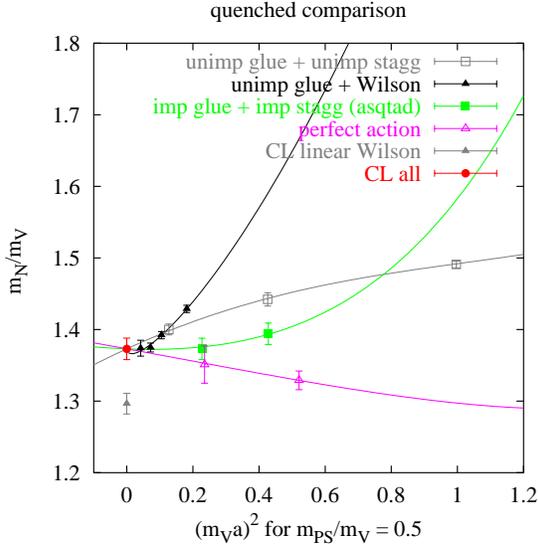}}
\caption[gjkjg]{$m_N/m_V$ for $m_{PS}/m_V=0.5$ for a variety of quark formalisms 
in the quenched approximation. The curves are for a fit, described in the 
text, with a single continuum limit, marked with a filled circle. 
The filled grey triangle at $a$=0
shows the previous continuum limit for Wilson quarks obtained with a purely 
linear fit.}
\label{fig:aokisq}
\end{figure}

\section{NEW ANALYSIS AND RESULTS}

Figure~\ref{fig:aokisq} shows an updated plot of the Wilson and unimproved staggered 
results from Lattice 2000, along with new results from improved 
staggered (asqtad) quarks on improved glue~\cite{newmilc} 
and the perfect action~\cite{perfect}.
The new results were chosen for small statistical errors, 
being available at the quark mass required here (with interpolation), 
for more than one value of the lattice spacing and with a reasonably large
physical volume.  Note that the $x$ axis is now $a^2$ instead of $a$. 

The lines represent a simultaneous fit to all the data with 
a single continuum limit imposed. A good $\chi^2$ is obtained. 
The fit allows for leading and higher order polynomials 
in $a$ appropriate to each formulation (i.e even powers for all except 
Wilson) with a constraint placed upon the coefficients to avoid 
losing control of the fit. The effectiveness of this Bayesian approach~\cite{lepage} has been widely demonstrated for fitting correlators, chiral 
extrapolations etc.  -- all situations in which a critical issue is 
the systematic error in the final result 
which arises from leaving out higher order terms 
in an expansion. In all these cases we have a good physical understanding of 
the expansion and can therefore place constraints on the higher order 
terms. Taking the continuum limit is just such a case. We expect 
the scale of discretisation errors to be set by the size of internal 
momenta inside the hadron. This should be roughly a few hundred MeV. 
Here the priors on the coefficients in the polynomial in $m_Va$ were 
taken to be $\pm$0.5, and 5 terms were included in the expansion for 
each formulation (a constant plus 4 appropriate powers).

\begin{figure}[t]
\vspace{9pt}
\rotatebox{270}{\includegraphics[width=\hsize]{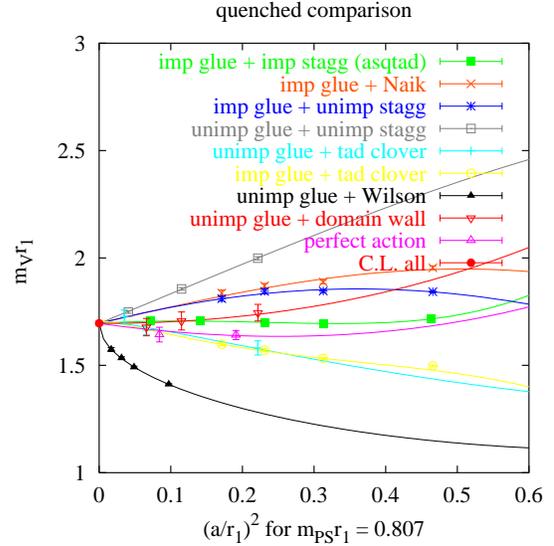}}
\caption{$m_V$ in units of $r_1$ at $m_{PS}r_1$=0.807. Results 
are from different quark formalisms and the fit has a joint 
continuum limit, as described in the text.} 
\label{fig:rho}
\end{figure}

The result for the joint continuum limit is 1.373(10), 
2$\sigma$ below
the previous unimproved staggered continuum limit but well above 
the Wilson one. With hindsight 
this is not surprising because a purely linear fit to Wilson 
results neglects terms which are not small for these results. We would 
expect the quadratic terms to appear as, say, $(m_Va/2)^2$ and this 
is 5\% for the coarsest lattice used. A systematic error has to be 
added to a purely linear extrapolation to take this into account. 
For the improved formalisms this is not so much of an issue because 
the terms neglected by a leading order extrapolation are $\cal{O}$$(m_Va/2)^4$, 
1\% for the improved staggered and perfect actions at their 
coarsest spacing. In all cases, however, a Bayesian analysis allowing for 
higher order terms is useful in assessing the systematic errors.  

\begin{figure}[t]
\vspace{9pt}
\rotatebox{270}{\includegraphics[width=\hsize]{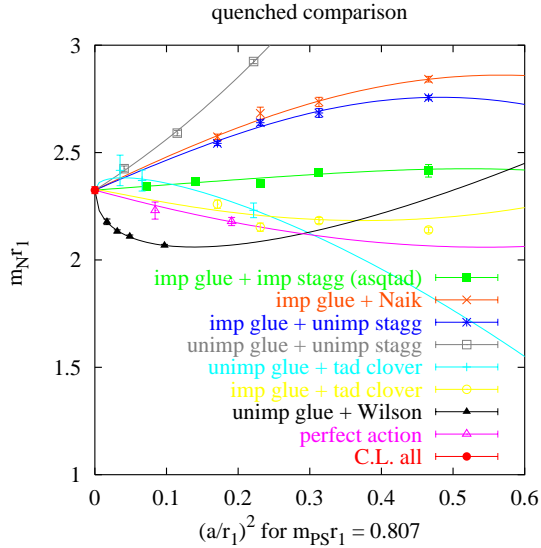}}
\caption{$m_N$ in units of $r_1$ at $m_{PS}r_1$=0.807. Results 
are from different quark formalisms and the fit has a joint 
continuum limit, as described in the text.} 
\label{fig:nuc}
\end{figure}

We now turn to other scaling plots, Figures~\ref{fig:rho} and~\ref{fig:nuc}, for which a lot more data 
is available. These are plots of $m_N$ and $m_V$ in units 
of $r_1$, vs $(a/r_1)^2$~\cite{doug}. We include results from Wilson~\cite{CP-PACS}, 
clover~\cite{collins,UKQCD}, perfect action~\cite{perfect}
and domain wall quarks~\cite{blum} as well as a variety of staggered 
results with improved and unimproved glue. Again a good fit with a single 
continuum limit is readily obtained and the curves for this are shown. 
The results are for a fixed quark mass given by $m_{PS}r_0=1.127$ using 
$r_0/r_1 = 1.397$ in the quenched continuum limit~\cite{newmilc}.  

The authors would welcome additional results for any of these graphs. 

\section{CONCLUSION}

We show, reassuringly, that all current quark formalisms give 
the same answer for nucleon and rho masses, at fixed arbitrary quark mass, 
in the quenched continuum limit. We used an analysis of discretisation effects which allows 
the inclusion of leading {\it and} higher order terms in the lattice 
spacing. This is particularly necessary for unimproved actions, 
emphasising once again the importance of the improvement programme in 
reducing systematic errors in lattice calculations. 

\vspace{3mm}
{\bf Acknowledgements}

We are grateful to Sinya Aoki, Peter Hasenfratz and Karl Jansen for useful discussions and 
to all the collaborations who provided results. 
This research was supported by PPARC, the EU, DoE and NSF.

\end{document}